\documentclass[amssymb,12pt,showpacs,aps]{revtex4}
\usepackage{graphicx}
\begin{document}
\title{Complexity of spectral sequences: semiclassical approach}
\date{\today}
\author{Yu. Dabaghian}
\affiliation{Department of Physiology,
Keck Center for Integrative Neuroscience,\\
University of California,
San Francisco, California 94143-0444, USA, \\ 
e-mail: yura@phy.ucsf.edu}

\begin{abstract}
It has been long recognized that the task of semiclassical evaluation of 
quantum spectra for the classically nonintegrable systems is fundamentally 
more complex than for the classically integrable ones. Below it is argued 
that the quantum spectra of the chaotic systems can differ among themselves
by level of their complexity. 
\end{abstract}
%\pacs{17}
\maketitle

\section{Introduction}
\label{section:introduction}

It is well known that the structure of the spectrum of a quantum mechanical
system reflects the dynamical properties of its classical counterpart. The
degree of classical dynamical regularity is reflected in the general
statistical properties of the quantum spectra and in the nature of the
analytical solution of the spectral problem.

For the conservative systems, dynamical regularity is usually understood as
the possibility to specify the trajectories via a set of quantities that
remain constant throughout dynamical evolution -- the integrals of motion.
If each one of the $d$ degrees of freedom of a bounded system corresponds to
a conserved quantity, $I_{1}$, ..., $I_{d}$, then such a system is completely 
integrable and its dynamical behavior can be regarded, in suitable coordinates, 
as a combination of oscillations. Algebraically, this is a manifestation of the 
underlying symmetries in the system, that lead to the appearance of certain global 
structures in the phase space known as integrable tori \cite{Arnold}, that uphold 
the oscillatory pattern of the trajectories.

As it was pointed out by the EBK theory, in semiclassical regime the integrals 
of motion assume a set of discrete values, defined via the so called quantum
numbers, $n_{i}$, 
\begin{equation}
I_{i}\sim \hbar \left(n_{i}+\frac{\mu_{i}}{4}\right),
\label{ints}
\end{equation}
where $\hbar$ is the Plank's constant and $\mu_{i}$ is the Maslov index
\cite{Gutzw,Cvit}. In view of (\ref{ints}), each integral of motion provides a
uniform global map from the natural numbers into the spectral sequence, which 
leads to the analytical solution to the corresponding quantum mechanical spectral 
problem. For example, if the energy is defined in terms of the motion integrals 
as $E=E^{(\mathop{\rm class})}\left(I_{1},...,I_{d}\right)$, the quantum eigenvalues 
of energy are
\begin{equation}
E^{(\mathop{\rm class})}\left( I_{1},...,I_{d}\right) \rightarrow 
E^{(\mathop{\rm quant})}\left(\hbar \left( n_{1}+\frac{\mu_{1}}{4}\right) 
,...,\hbar \left( n_{d}+\frac{\mu_{d}}{4}\right) \right).
\label{EBK}
\end{equation} 
The integers $n_{i}$ can be interpreted as the number of times the wave 
representing the quantum mechanical particle winds around the basic torus 
cycles, so physically, the discretization of the spectrum in the EBK 
theory can be understood as a manifestation of the geometrical consistency 
between the (semiclassical) quantum waves and the dynamical trajectories.

In contrast, if the number of motion integrals is less than the number of the 
degrees of freedom, then the dynamics is ``irregular'', in the sense that its 
trajectories do not follow any particular patterns in the phase space. Rather, 
they tend to cover as much phase space $X$ as possible at a given energy, more 
or less uniformly, so the only relevant long term characteristics of the dynamics 
are the relative frequencies of the visitations of different parts of $X$, which 
can be equally well described by a certain smooth probability distribution $P(x)$.

As a result, the complexity of the semiclassical quantization task for the
nonintegrable systems differs substantially from the integrable ones. While
the EBK quantization method is constructive, i.e. it is possible to define 
the {\em individual} levels of the quantized integrable systems explicitly, 
by specifying a particular set of quantum numbers (\ref{EBK}), for nonintegrable 
systems such individualized solution for the spectral problem is generally
unavailable. Instead, one of the main results of the semiclassical theory is
the series expansion representation for the quantum density of states, which
is a {\em global} characteristics of the spectrum, the so-called
Gutzwiller's formula: 
\begin{equation}
\rho (E) =\sum_{n}\delta \left( E-E_{n}\right) =\bar{\rho}(E) +
\mathop{\rm Re}\sum_{p}B_{p}(E) e^{iS_{p}(E)}.
\label{Gutzw}
\end{equation}
Here $S_{p}(E) $ is the action functional evaluated for the periodic orbit $p$, 
and $B_{p}$ is a certain weight factor \cite{Gutzw}.

One can immediately appreciate the difference in complexity of the series 
(\ref{Gutzw}) compared to the (\ref{EBK}). The sum (\ref{Gutzw}) includes 
all the periodic orbits (that are isolated in a fully chaotic system), and 
so it reflects the full complexity of the classical phase space structure 
of a nonintegrable system. The essence of the Gutzwiller's formula is that 
the oscillating amplitudes $e^{iS_{p}(E)}$ produced by the periodic orbits, 
combined with the appropriate weights $B_{p}$, produce a constructive 
interference effect every time $E$ happens to be equal to one of the quantum 
eigenvalues and cancel each other out for all other values of $E$, which enables 
one to transform the information contained phase space structures into the 
pattern of quantum mechanical spectral sequence \cite{Gutzw}.

An important aspect of (\ref{Gutzw}) is that it is indeed a very general 
result, that can be applied to a great variety of systems. Moreover, certain 
mathematical systems or objects can be put into a quantum chaotic context 
{\em because} their constituents can be related via formula (\ref{Gutzw}) 
and hence be interpreted as ``quantum chaotic''. One well known example of 
this is provided by the relationship between the eigenvalues of the Laplace 
- Beltrami operator on a surface of constant negative curvature and the 
dynamics of a point particle moving on it. This connection is described by 
the famous Selberg trace formula \cite{Selberg}, that was first discovered 
as a number theoretic and functional analysis theorem. 
Apart from the deterministic chaotic systems, the expansion (\ref{Gutzw})
can also describe the spectra of many classically stochastic systems, e.g. 
the so-called quantum graphs \cite{QGT,Gaspard} and 2D ray splitting billiards
\cite{Blumel}: as long as the stochastic dynamics generates trajectory patterns 
similar to the deterministic chaotic trajectories, they manifest themselves in 
the in the same physical context in quantum regime.

Perhaps the most intriguing example is provided by the relationship between 
the zeroes of Riemann's zeta function and the set of the prime numbers. In 
\cite{Berry1,Berry3,Berry4} it was argued that the integrated spectral density, 
the spectral staircase of the imaginary parts of the nontrivial zeroes of 
Riemann's zeta function, $E_{n}=\mathop{\rm Im}(z_{n})$, $\zeta(z_{n}) =0$,
\begin{equation}
N(E) =\sum_{n}\Theta\left(E-E_{n}\right) 
\label{Stair}
\end{equation}
can be expanded into the Gutzwiller
type series (\ref{Gutzw}) in which the average part of the density of 
$E_{n}$s is given by
\begin{equation}
\bar{N}(E) =\frac{E}{2\pi}\left( \ln \left( \frac{E}{2\pi}\right)-1\right) 
+\frac{7}{8}+...,
\label{WeylRiemann}
\end{equation}
and the oscillating part by
\begin{equation}
\delta N(E) \approx -\frac{1}{\pi}\mathop{\rm Im}\sum_{p}
\sum_{m=1}^{\infty}\frac{e^{imE\ln p}}{mp^{\frac{m}{2}}}, 
\end{equation}
where the index $p$ runs over the set of prime numbers. Hence the prime
numbers in this case play the role of the prime periodic orbits of lengths
$L_{p}=\ln p$. This shows that the organization of the roots of the zeta 
function, which a is purely number theoretic object, is as irregular with 
respect to the set of prime numbers, as is the quantum spectrum of a 
classically chaotic system as expressed by the ``periodic orbit expansion'' 
\cite{Bogomolny2,Cvit}.
%%%%%%%%%%%%%%%%%%%%%%%%%%%%%%%%%%%%%%%%%%%%%%%%%%%%%%%%%%%%%%%%%%%%%%
\begin{figure}[tbp]
\begin{center}
\includegraphics{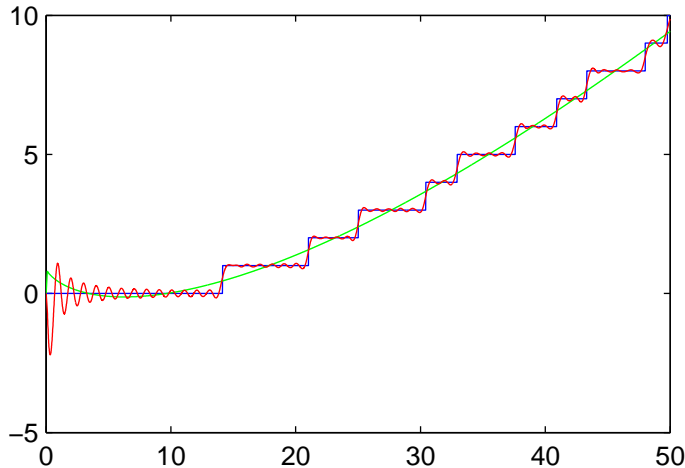}
\end{center}
\caption{Spectral staircase for the imaginary parts of zeroes of
Riemann's zeta function. Wiggly line represents the Gutzwiller's series 
for $N(k)$, and the smooth line running through the staircase is the
Weyl's average $\bar{N}(k)$.}
\label{riemann_stair}
\end{figure}
%%%%%%%%%%%%%%%%%%%%%%%%%%%%%%%%%%%%%%%%%%%%%%%%%%%%%%%%%%%%%%%%%%%%%%
Despite these significant implications of Gutzwiller's formula, it does not
produce the final solution of the spectral problem in the same sense as the
EBK theory does for the integrable systems. The problem is that the
Gutzwiller's formula does not specify the {\em individual} energy or
momentum eigenvalues, i.e. it does not produce a functional correspondence
between the spectral sequence and the natural number (quantum number)
sequence, although such correspondence actually can exist.

In order to evaluate the individual eigenvalues $E_{n}$ from (\ref{Gutzw}),
one needs to provide additional {\em local} information, that would allow
to single out the individual delta peaks in $\rho(E)$. To have a uniform 
solution to the spectral problem that would be equivalent to the EBK formula 
(\ref{EBK}), one must be able to produce such information systematically, 
in a way that would eventually lead to an explicit functional dependence 
$E_{n}=E(n)$.

It is easy to show that in principle this not an impossible task. Indeed,
consider a the spectral staircase (\ref{Stair}), that monotonously increases 
with energy. Consider also a smooth monotone function $f(E) $ that intersects 
every stair step of $N(E)$ as shown on Fig. 2.
%%%%%%%%%%%%%%%%%%%%%%%%%%%%%%%%%%%%%%%%%%%%%%%%%%%%%%%%%%%%%%%%%%%%%%
\begin{figure}[tbp]
\begin{center}
\includegraphics{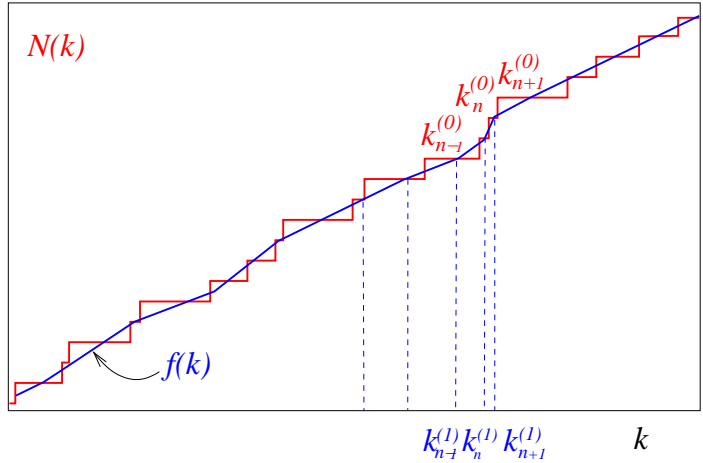}
\end{center}
\caption{Spectral staircase $N(k)$ for the eigenvalues of momenta of quasi one 
dimensional quantum networks (quantum graphs), that are known to be stochastic
in classical limit (see. Section \ref{section:graphs}). A smooth monotone
function $f(k)$ intersects every stair-step of $N(k)$.}
\label{function}
\end{figure}
%%%%%%%%%%%%%%%%%%%%%%%%%%%%%%%%%%%%%%%%%%%%%%%%%%%%%%%%%%%%%%%%%%%%%%
It is also possible to chose $f(E) $ in such a way that the average deviation 
of $f(E) $ from $N(E) $ vanishes, $\left\langle f(E) -N(E) \right\rangle =0$, 
so $f(E) $ can be considered as an ``average'' of $N(E) $, that is in general 
{\em different} from the Weyl's' average \cite{Baltes}. The intersection points 
$\hat{E}_{n}$, $f\left( \hat{E}_{n}\right) =N\left( \hat{E}_{n}\right) =n$, 
satisfy the condition
\begin{equation}
...\leqslant E_{n-1}\leqslant \hat{E}_{n}\leqslant E_{n}\leqslant \hat{E}
_{n+1}\leqslant E_{n+1}\leqslant ...\ .  
\label{sepE}
\end{equation}
Since $f(E) $ is monotone, it can be inverted, so the roots of 
$f(\hat{E}_{n})=N\left( \hat{E}_{n}\right) $, can be defined as
a function of $n$, $\hat{E}_{n}=f^{-1}(n) =\hat{E}(n)$. The resulting 
separating sequence (\ref{sepE}) can then be used to find the eigenvalues 
$E_{n}$ in the form of an explicit formula \cite{Opus,Prima,Sutra,Stanza},
\begin{equation}
E_{n}=E(n) =\int_{\hat{E}(n-1)}^{\hat{E}(n)}\rho(E') E'dE'.
\label{regE}
\end{equation}
Due to the expansion (\ref{Gutzw}), one can in principle evaluate the
integral (\ref{regE}) explicitly and obtain the formula for $E_{n}=E(n)$ 
in an expansion form structurally similar to (\ref{Gutzw}). In general, 
the structure of the classical phase space and hence the structure of the 
sum (\ref{Gutzw}) can change in very complex ways as a function of energy, 
which makes the evaluation of the integral (\ref{regE}) a much more
complicated problem. In order to avoid these difficulties, the following
discussion will be restricted to the so-called scaling systems, e.g. the
billiards, in which case the expansion coefficients $B_{p}$ are energy
independent and the functional structure of the periodic orbit sum (\ref
{Gutzw}) is fixed.

Such construction produces a map from the natural number sequence first 
into the separating sequence $\hat{E}_{n}=f^{-1}(n)$, and then to physical 
spectral sequence $E_{n}$ via (\ref{regE}). Hence in fully nonintegrable 
system there also exists a solution to the spectral problem that produces 
the eigenvalues $E_{n}$ as a function of $n$, $E_{n}=E(n)$ based on the 
semiclassical expansion (\ref{Gutzw}).

Formulae (\ref{sepE}) and (\ref{regE}) suggest that the possibility to solve
the spectral problem depends on the possibility to localize the individual
delta peaks within intervals $I_{n}=\left[ \hat{E}_{n},\hat{E}_{n-1}\right]$, 
loosely defined by the inequality (\ref{sepE}). From such perspective, the
missing part of the solution is the information about the structural
complexity if the spectral sequence.

\section{Structural analysis}
\label{section:structural}

The task of obtaining $\hat{E}_{n}$ is in fact much more complicated than 
it may appear at the first sight. Finding a monotone function that would 
follow the pattern of $N(k) $ in the detail required by (\ref{sepE}) for 
a generic system proves to be an extremely difficult problem \cite{Baltes}. 
As a result, outside of a few simple systems 
\cite{Opus,Prima,Sutra,Stanza,Saga,Anima,Fabula}, there are virtually no 
examples of explicit functional dependencies of the quantum energy levels,
$E_{n}$, on the index $n$, $E=E(n)$, for the classically nonintegrable 
systems.

The apparent impenetrability of the spectral problem for quantum chaotic
systems seems to impose an implicit (and false) empirical dichotomy --
either the system is integrable and explicitly quantizable via EBK theory,
or it is nonintegrable and the spectral problem does not have an explicit
solution. However, as it was argued above, even for a completely
nonintegrable systems it is still possible to find the semiclassical solution
to the spectral problem, $E_{n}=E(n)$. Moreover, the intricacy of the quantum 
chaotic spectra may conceal a rich complexity structure, that can be studied 
both within the paradigm of semiclassical physics and outside of it.

In itself, the mathematical problem is to describe the complexity of mapping
the natural numbers $n=1$, $2$, ... into the spectral sequence $E_{n}=E(n)$. 
However, in the specific context of quantum chaos theory, the task is to do 
this within the paradigm of the semiclassical physics, in which the objects 
and the phenomena of the classical dynamics provide the semantics for describing 
quantum objects and phenomena. The question is then, how uniform the description 
of spectra may be from the point of view of algorithmic complexity?

It is clear that {\em a priori}, the regularity of the delta peak patterns in 
$\rho(k)$ may be different for different quantum chaotic systems, so the effort 
required for establishing a mapping from the natural numbers into the the sequence 
of $\hat{E}_{n}$'s (and hence into the $E_{n}$'s) may be different, so the 
semiclassical solutions of spectral problems for the nonintegrable systems may 
not be algorithmically equivalent.

Let us examine the regularity of the spectral sequences more closely. Let 
$\bar{N}(E)$ be the average number of the spectral points on the interval
$[0,E]$, so that 
$N(E)=\bar{N}(E)+\delta N(E)$, $\left\langle \delta N(E)\right\rangle=0$. 
For example, $\bar{N}(E) $ can be the Weyl's average, given by the volume of 
the phase space of the system for the corresponding $E$. This function can 
be used to define the unfolded spectral sequence 
\begin{equation}
k_{n}^{(0)}=\bar{N}(E_{n}), 
\end{equation}
distributed with uniform average density $\bar{\rho}^{(0)}(k) =1$, so that 
\begin{equation}
k_{n}^{(0)}=n+\delta_{n}^{(0)} ,
\label{unfolded}
\end{equation}
with $\left\langle \delta_{n}^{(0)}\right\rangle =0$. This
unfolding operation provides a common ground for studying spectral sequences
coming from different systems.

Clearly, the possibility to find an analytically defined bootstrapping sequence 
depends on the magnitude of the fluctuations $\delta_{n}^{(0)}$. In the simplest 
case, if the disorder of the original sequence $k_{n}^{(0)}$ is weak, so that it 
is sufficiently close to a periodic sequence, then the periodic points
\begin{equation}
k_{n}^{(1)}=k^{(\mathop{\rm reg})}=n+\gamma^{(\mathop{\rm reg})},
\label{periodic}
\end{equation}
where $\gamma ^{(\mathop{\rm reg})}$ is a constant, can be interlaced with it 
according to (\ref{sepE}). Geometrically, this case corresponds to the situation 
when the spectral staircase $N(k)$ of the unfolded sequence can be pierced by 
the straight line average $\bar{N}(k)$. 
%%%%%%%%%%%%%%%%%%%%%%%%%%%%%%%%%%%%%%%%%%%%%%%%%%%%%%%%%%%%%%%%%%%%%%
\begin{figure}[tbp]
\begin{center}
\includegraphics{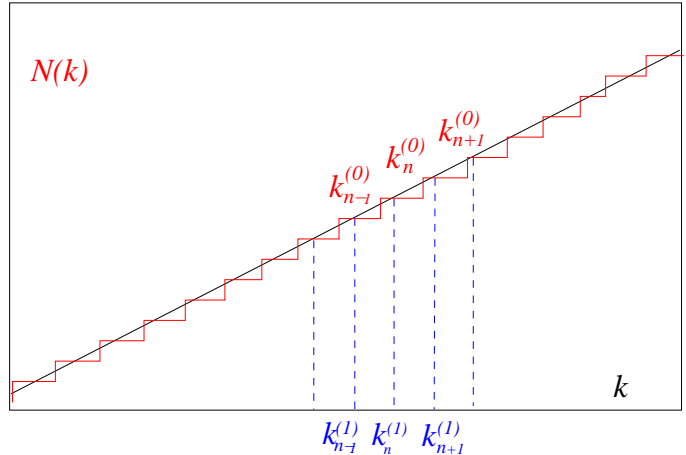}
\end{center}
\caption{Spectral staircase for unfolded regular spectral sequence. Every 
stair-step of $N(k)$ is pierced by a single line.}
\label{reg_stair}
\end{figure}
%%%%%%%%%%%%%%%%%%%%%%%%%%%%%%%%%%%%%%%%%%%%%%%%%%%%%%%%%%%%%%%%%%%%%%
The systems with such ``almost integrable'' spectral behavior were 
referred to as {\em regular} in \cite{Opus,Prima,Sutra,Stanza}. The 
numerical values for $k_{n}^{(0)}$ in this case can be computed explicitly 
given the explicit expansion of $\rho(k)$, via
\begin{equation}
k_{n}^{(0)}=\int_{n-1+\gamma^{(1)}}^{n+\gamma^{(1)}}k\rho^{(0)}(k)dk.
\label{regn}
\end{equation}
It is clear that a priori, the stair steps of a generic sequence's staircase 
can not all be pierced by a single straight line, so a single periodic sequence 
that bootstraps the spectrum does not exist. In other words, a generic sequence 
$k_{n}^{(0)}$ is so disordered, that any sequence $k_{n}^{(1)}$ that bootstraps 
it cannot itself be periodic. Instead, the bootstrapping requires a smooth 
monotone curve $f(k)$ that intersects each stair-step of the spectral staircase 
$N(k)$ and generates a certain aperiodic sequence $k_{n}^{(1)}$ that is 
circumscribed in $k_{n}^{(0)}$, as shown on Fig. \ref{function}. 

On the other hand, it is clear that 
if the bootstrapping function $f(k)$ envelops the staircase $N(k)$ maximally 
smoothly and uniformly, so that $\left\langle f(k)-N(k)\right\rangle=0$, then the 
sequence $k_{n}^{(1)}$ is more ordered than $k_{n}^{(0)}$. Since $k_{n}^{(1)}$ is 
more ordered than $k_{n}^{(0)}$, it may happen that $k_{n}^{(1)}$ {\em itself} can 
be bootstrapped by a periodic sequence, in which case $k_{n}^{(0)}$ will be 
bootstrapped with the periodic sequence 
$k_{n}^{(\mathop{\rm reg})}=n+\gamma^{(\mathop{\rm reg})}$ in two steps, via one 
auxiliary sequence $k_{n}^{(1)}$. If however, the $k_{n}^{(2)}$ sequence that 
bootstraps $k_{n}^{(1)}$ must necessarily be aperiodic, then the question will be 
whether the sequence $k_{n}^{(3)}$ can be chosen periodic, an so on.

This immediately suggests a clear strategy of ``unfolding'' any sequence 
$k_{n}^{(0)}$ using an auxiliary set of bootstrapped sequences $k_{n}^{(1)}$, 
$k_{n}^{(2)}$, ..., $k_{n}^{(\mathop{\rm reg})}$, 
\begin{eqnarray}
k_{n}^{(0)} &\leqslant &k_{n}^{(1)}\leqslant k_{n+1}^{(0)},  
\label{bootstrap} \\
k_{n}^{(1)} &\leqslant &k_{n}^{(2)}\leqslant k_{n+1}^{(1)}  
\nonumber \\
&&...  
\nonumber \\ 
k_{n}^{(r-1)} &\leqslant &k_{n}^{(r)}\leqslant k_{n+1}^{(r-1)},  
\nonumber
\end{eqnarray}
that starts with the original sequence $k_{n}^{(0)}$ and terminates when 
the last sequence $k_{n}^{(r)}$ can be interlaced by a periodic sequence 
(\ref{periodic}), $...\leqslant n-1+\gamma^{(\mathop{\rm reg})}\leqslant 
k_{n}^{(r)}\leqslant n+\gamma^{(\mathop{\rm reg})}\leqslant k_{n+1}^{(r)}
\leqslant n+1+\gamma ^{\left( reg\right)}\leqslant ...\ $. At each step, 
the auxiliary sequences are chosen in such a way that the size of the 
fluctuations $\delta_{n}^{(j+1)}=\left( k_{n}^{(j)}-n\right)$ decreases 
with the increase of the index $j$, so starting from the original
sequence $k_{n}^{(0)}$, each following sequence is closer to periodic.
%%%%%%%%%%%%%%%%%%%%%%%%%%%%%%%%%%%%%%%%%%%%%%%%%%%%%%%%%%%%%%%%%%%%%%
\begin{figure}[tbp]
\begin{center}
\includegraphics{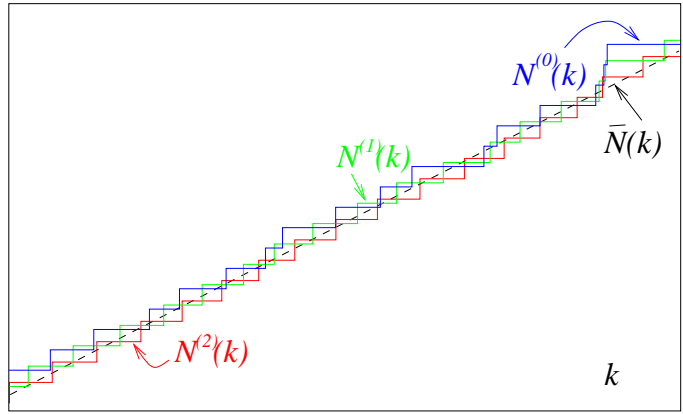}
\end{center}
\caption{Bootstrapped spectral staircase $N^{(0)}(k)$s and 2 auxiliary  
sequences, $N^{(1)}(k)$ and $N^{(2)}(k)$ obtained for the quantum momentum 
spectrum of a fully connected 4-vertex quantum network. The regular staircase
is pierced by the line $\bar N(k)$.}
\label{Bootstrapping}
\end{figure}
%%%%%%%%%%%%%%%%%%%%%%%%%%%%%%%%%%%%%%%%%%%%%%%%%%%%%%%%%%%%%%%%%%%%%%
Let $\rho^{(j)}(k)$ be the density of the separating points at the level 
$j$, so $N^{(j)}(k)$ is the spectral staircase for the points $k_{n}^{(j)}$. 
Then the values in two bootstrapped sequences can be related to one another
according to 
\begin{equation}
k_{n}^{(j-1)}=\int_{k_{n-1}^{(j)}}^{k_{n}^{(j)}} k'dN^{(j-1)}(k').
\label{levelj}
\end{equation}
If the harmonic expansion similar to (\ref{Gutzw}) is known for each
sequence $k_{n}^{(j)}$, 
\begin{equation}
N^{(j)}(k) =\bar{N}(k) +\mathop{\rm Im}
\sum_{p}A_{p}^{(j)}e^{iS_{p}^{(j)}(k)},
\label{Nj}
\end{equation}
with explicitly defined $A_{p}^{(j)}$ and $S_{p}^{(j)}(k)$ 
\cite{QGT,Nova,HBohr,Corduneanu,Besicovitch}, then the
integral (\ref{levelj})\ can be evaluated explicitly, and so the sequence 
$k_{n}^{(j)}$ is explicitly mapped onto $k_{n}^{(j-1)}$. Due to the mappings 
(\ref{levelj}) the index $n$ propagates through the hierarchy of bootstrapped 
sequences (\ref{bootstrap}) and emerges in the $0$th sequence $k_{n}^{(0)}$. 
The number of auxiliary sequences required for bootstrapping $k_{n}^{(0)}$ 
with $k_{n}^{(r)}$ therefore produces a certain complexity index for the 
quantum spectra that shows how a global index $n$ can be consistently mapped
onto the spectral sequence of any degree of disorder.

There are certainly many ways in which a given sequence $k_{n}^{(j)}$ can be 
bootstrapped. The number of levels in the hierarchy depends on the choice of 
the algorithm for obtaining the bootstrapping sequences $k_{n}^{(j)}$. Given 
a strategy for generating the bootstrapping sequences, the regularity of the 
sequence can be determined empirically, either by parsing through its individual 
elements or by studying the probability of occurrence of the fluctuation magnitudes. 
In general, a separating sequence for $k_{n}^{(j)}$ can be written as 
\begin{equation}
k_{n}^{(j+1)}=
\alpha_{n}^{(j)}k_{n+1}^{(j)}+\left(1-\alpha_{n}^{(j)}\right) k_{n}^{(j)},
\label{separating}
\end{equation}
where $0<\alpha_{n}^{(j)}<1$ are arbitrary
parameters. To minimize the deviation from a periodic sequence, one can consider 
the functional
\begin{equation}
F=\sum\limits_{n}\left(\left(k_{n}^{(j+1)}-k_{n-1}^{(j+1)}\right)-T\right)^{2}
=\sum\limits_{n}\left( \left( \alpha_{n}^{(j)}s_{n}^{(j)}+\left(1-\alpha_{n-1}^{(j)}
\right) s_{n-1}^{(j)}\right)-T\right) ^{2},  
\label{functional}
\end{equation}
where, for the fully unfolded sequences, $s_{n}^{(j)}=\left(1+\delta_{n+1}^{(j)}-
\delta_{n}^{(j)}\right)$ and $T=1$. Varying $F$ with 
respect to parameters $\alpha_{n}^{(j)}$ under the conditions 
$0\leqslant\alpha_{n}^{(j)}\leqslant 1$ and $k_{n}^{(j+1)}>k_{n-1}^{(j+1)}$, 
one gets 
\begin{equation}
\alpha_{n}^{(j)}=\frac{1}{s_{n}^{(j)}}
\left( \delta_{0}^{(j)}-\delta_{n}^{(j)}\right)
\label{optimal}
\end{equation}
if $0\leqslant\alpha_{n}^{(j)}\leqslant 1$, and $\alpha_{n}^{(j)}=0$ 
if $\alpha_{n}^{(j)}\leqslant 0$ and $\alpha_{n}^{(j)}=1$ if 
$\alpha_{n}^{(j)}\geqslant 1$, so that the optimal bootstrapping can 
be explicitly defined in terms of the original sequence, which is 
important for practical studies of empirically obtained sequences, e.g.
for studying the the sequences of numerically obtained eigenvalues of 
Schr\"odinger's operator. For crude estimates, other choices of 
$\alpha_{n}^{(j)}$'s can be used, e.g. $\alpha_{n}^{(j)}=1/2$ for 
$j=1$, ..., $r$. 

The suppression of the fluctuations across the sequences $k_{n+1}^{(j)}$ 
is also reflected in the statistical properties of the deviations $\delta_{n}^{(j)}$.
%%%%%%%%%%%%%%%%%%%%%%%%%%%%%%%%%%%%%%%%%%%%%%%%%%%%%%%%%%%%%%%%%%%%%%
\begin{figure}[tbp]
\begin{center}
\includegraphics{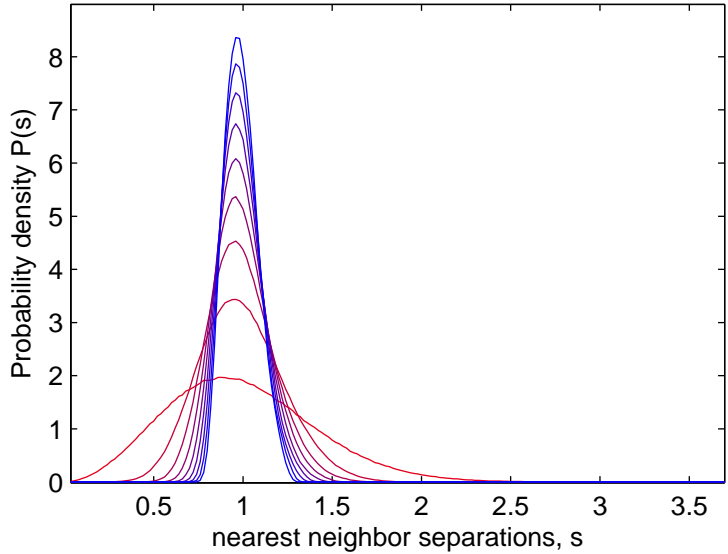}
\end{center}
\caption{The histogram of the nearest neighbor separations 
$s_{n}=\frac{\gamma_{n+1}-\gamma_{n}}{2\pi}\ln\left(\frac{\gamma_{n}}{2\pi}\right)$ 
for 100,000 roots of the Riemann's zeta function (found at A. Odlyzko's 
website, 
$\mathop{\rm http://www.dtc.umn.edu/~odlyzko/zeta\mathop \_ tables}$) 
and the separating sequences obtained using $\alpha_{n}^{(j)}=1/2$. 
Since the the first sequence in which the nearest neighbor separation 
magnitude does not exceed 1 it shows that the irregularity of Riemann 
zeros is less than 7. If the solutions (\ref{optimal}) are used, then 
the complexity degree of the resulting hierarchy is 2.}
\label{Separations}
\end{figure}
%%%%%%%%%%%%%%%%%%%%%%%%%%%%%%%%%%%%%%%%%%%%%%%%%%%%%%%%%%%%%%%%%%%%%%
The results of numerical analysis of the spectral fluctuations of 
quantum chaotic systems show that the probability of having large
fluctuations decreases with the increase of hierarchy index $j$ 
(Fig~\ref{Separations}). 
Although at the $0$th (physical) level of the hierarchy the histograms 
of various spectral statistics show the characteristic universal 
features \cite{BGS,ABS}, the shape of the distributions at the higher 
levels may deviate from them. Numerical studies indicate that the 
distribution of various spectral characteristics at the regular level 
(especially for the hierarchies with high $r$) tends to have a Gaussian-like 
shape, which develops through the hierarchy into different physical 
distribution profiles. For example, for the nearest neighbor statistics, 
$s_{n}^{(j)}=\left(1+\delta_{n+1}^{(j)}-\delta_{n}^{(j)}\right)$, it
develops into Wignerian-like distribution, or, for $\delta_{n}^{(j)}$ 
or $\xi_{n}^{(j)}=\left(\delta_{n+1}^{(j)}+\delta_{n}^{(j)}\right)/2$ 
it develops into a Gaussian distribution with a larger variance, etc. 
A typical illustration of the appearance of the universal probability 
distribution profiles at the most disordered, $0$th, level of the 
hierarchy out of the distributions of the the regular, $r$th, level is 
shown on Fig.~\ref{Separations}.
Clearly, the support of the probability distribution functions becomes 
progressively wider with the approach to the physical level of the 
hierarchy. At the regular level, the range of the probability distribution 
function is defined by the condition $|\delta_{n}^{(j)}|<1$.

\section{Complexity of quantum graph spectra}
\label{section:graphs}

\begin{figure}[tbp]
\begin{center}
\includegraphics{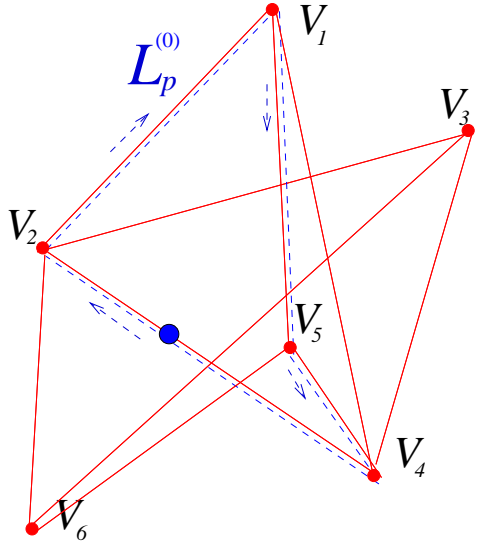}
\end{center}
\caption{A 6 vertex quantum graph. Dashed line represents a periodic 
trajectory $p$ with the length $L^{(0)}_{p}$.}
\label{graph}
\end{figure}

In \cite{Opus,Prima,Sutra,Stanza,Anima,Fabula,Saga} it was shown that this
approach can be applied to the case of the quantum graphs -- simple quasi 
one dimensional, scaling, classically stochastic models (Fig. \ref{graph}), 
that are often used to model deterministic chaotic behavior in low dimensional 
dynamical systems \cite{QGT,Gaspard}.
%%%%%%%%%%%%%%%%%%%%%%%%%%%%%%%%%%%%%%%%%%%%%%%%%%%%%%%%%%%%%%%%%%%%%%
%%%%%%%%%%%%%%%%%%%%%%%%%%%%%%%%%%%%%%%%%%%%%%%%%%%%%%%%%%%%%%%%%%%%%%
In \cite{Anima,Saga,Fabula} it was shown that there exist quantum graphs 
with different degrees of spectral irregularity defined in the sense of 
the bootstrapping hierarchy. For the simplest case of the regular quantum 
graphs \cite{Opus,Prima,Sutra,Stanza,Wilmington}, the spectrum obtained 
from (\ref{regn}) is given by
\begin{equation}
k_{n}^{(0)}=\frac{\pi n}{L_{0}}+\mathop{\rm Im}\sum\limits_{p}C
_{p}^{(0)}e^{iL_{p}^{(0)}\frac{\pi n}{L_{0}}},
\label{kn0}
\end{equation}
where the coefficients $C_{p}^{(0)}$ emerge from the parameters 
of the Gutzwiller's series \cite{QGT,Nova} after the integration 
(\ref{regn}), $L_{0}$ is the total length of the graph bonds, and 
$L_{p}^{(0)}$s are the lengths of the periodic orbits $p$ on the graph.

The expansion (\ref{kn0})\ suggests a simple physical interpretation. 
As mentioned above, a quantum graph is regular, if its spectrum is 
sufficiently close to the periodic sequence 
\begin{equation}
k_{n}^{(r)}=\frac{\pi n}{L_{0}}.  
\label{int}
\end{equation}
Incidentally, (\ref{int}) defines the ``integrable'' spectrum of a 
point particle in a $1D$ box of the same overall size $L_{0}$. 
According to (\ref{kn0}), the required corrections to (\ref{int}) 
are given by the semiclassical amplitudes $e^{iS_{p}^{(r)}}$, where 
$S_{p}^{(r)}=L_{p}^{(0)}k_{n}^{(r)}$ are the values of the action 
functionals evaluated for all the available periodic motions in 
the system with the ``integrable'' momentum (\ref{int}). So the $n$th 
level quantum wave that is geometrically consistent with the graph's 
size, as it would be in the case infinite square well, now also 
circles simultaneously along all the possible periodic paths on the 
graph with the momentum (\ref{int}).
%, producing the ``nonintegrability corrections''. 
The resulting periodic orbit fluctuation amplitudes, combined with 
the right weights, produce the actual momentum eigenvalues for the 
nonintegrable system. Hence the exact ``nonintegrable'' values of 
$k_{n}^{(0)}$ also appear as the result of a complex interference 
effect similar to (\ref{Gutzw}).

This represents a generalization of the EBK quantization formula 
for the case of quantum graphs, that allows to build the individual 
eigenvalues $k_{n}$ constructively, via the explicit functional 
dependence on the index $n$. The effect of classical nonintegrability 
appears in (\ref{kn0}) as a ``perturbation'' to the integrable
background pattern, if $k_{n}-\frac{\pi n}{L_{0}}<\frac{\pi}{L_{0}}$. 
In this aspect, the situation is reminiscent of the effect of 
persistence of the integrability structures in the phase space 
under small nonintegrable perturbations described by the KAM theory 
\cite{Arnold}.

For the irregular graphs with several level spectral hierarchy, the
relationship (\ref{kn0}) is repeated for each transition between the
hierarchy levels, 
\begin{equation}
k_{n}^{(j)}=k_{n}^{(j+1)}+\eta ^{(j+1)}+
\mathop{\rm Im}\sum\limits_{p}C_{p}^{(j)}e^{iL_{p}^{(j)}k_{n}^{(j+1)}},
\label{knj}
\end{equation}
where $\eta^{(j+1)}$ is a bounded function of the fluctuations on the 
$j+1$ level, $\eta^{(j+1)}=\eta^{(j+1)}\left(\delta^{(j+1)}\right)$
\cite{Gratia,Tantra,Magna}. 
Also in general, the harmonic expansion (\ref{knj}) is different from 
the periodic orbit expansion of (\ref{kn0}), 
$C_{p}^{(j)}\neq C_{p}^{(0)}$ and $L_{p}^{(j)}\neq L_{p}^{(j)}$ 
\cite{Gratia,Tantra,Magna}. Hence the $j$th level oscillations 
transmit the discrete momentum values from the $(j+1)$th to the $j$th 
level of the hierarchy. As a result, the ``integrable spectrum'' 
$k_{n}^{(r)}$ that explicitly carries the quantum number $n$ is adjusted 
$r$ times according to the set of equations (\ref{knj}) until it is 
transmitted from the regular to the $0$th level of the hierarchy. In 
\cite{Anima,Saga,Fabula} it was sown that every graph is characterized 
by a finite irregularity degree.

It is important that a quantum graph of given topology can have different
degrees of spectral irregularity depending on the bond length and other
graph parameters \cite{Anima,Saga,Fabula}. Since the geometric complexity 
of the periodic orbit set is defined by the topology of the graph, it 
means that spectral irregularity, as a complexity measure, is not a trivial
reflection of the phase space complexity of the underlying classical system,
and provides a separate characterization of the complexity of quantum spectra.

The expansions (\ref{kn0}) and (\ref{knj}) can be used to describe
analytically the development of the universal probability distributions 
\cite{Gratia,Tantra,Magna}, illustrated on Fig.~\ref{Separations}.
 
\section{Discussion}
\label{section:discussion}

The task of quantifying the complexity of the map between the natural
numbers and a given sequence $x_{n}=x(n) $ is very general. This problem 
was recently considered in \cite{Arnold1,Arnold2} for 
the case of finite sequences over finite alphabets, which revealed a
remarkably complex organizational structure of these mathematical objects.
The complexity organization scheme developed in \cite{Arnold1,Arnold2}
is generated by the linear difference operator
\begin{equation}
\Delta x(n) =x(n) -x(n+1) =x'(n),
\label{arnold}
\end{equation}
that maps one sequence into another. The motivation for using the operator 
$\Delta $ is that in certain simple cases it restricts the complexity of the
symbolic sequences, i.e. it produces more ordered sequences out of less
ordered ones. For example, the constant sequences $x_{n}\equiv x_{n}^{(0)}=
\mathop{\rm const}$, are mapped into $x_{n}^{\prime}=\Delta x_{n}^{(0)}=0$, 
which is the ``simplest'' constant sequence. If $x_{n}$ is linear, 
$x_{n}=x_{n}^{(1)}=an+b$, then its image $x^{\prime}=\Delta x_{n}^{(1)}=x_{n}^{(0)}$ 
is a constant sequence, which is simpler than linear, so that $\Delta ^{2}x^{(1)}=0$,
and so on. In general, the functions for which $\Delta ^{l}x=0$ for some $l$
are the polynomials of degree $m<l$, $x=p^{(m)}\in P_{l}$. Intuitively, the higher 
is the degree of the polynomial, the more complex is the mapping $x_{n}=x(n) $.

For the case of finite sequences, one necessarily runs not only into the
polynomials, but also into more complex ``exponential'' functions. By 
definition, the function for which $\Delta ^{l}x(n) =x(n) $ is an exponential 
polynomial $e^{(q)}\in E_{l}$ of the order $q$ that divides $l$. In 
\cite{Arnold1,Arnold2} it was shown that any function $f$ can be characterized 
by a polynomial degree $\deg(f) =\deg \left(p^{(m)}\right) =m$ and an exponential 
order $\mathop{\rm ord}(f) =\mathop{\rm ord}\left(e^{(q)}\right) =q$, 
which allows to formalize the organization of complexity of the sequences.

By definition \cite{Arnold1,Arnold2}, a function $x_{n}$ is
more complex than $x_{n}^{\prime}$ if $\mathop{\rm ord}(x_{n}) >\mathop{\rm 
ord}\left( x_{n}^{\prime}\right) $. If two sequences have the same order, 
$\mathop{\rm ord}\left(x'_{n}\right) =\mathop{\rm ord}\left(x'_{n}\right)$ 
then $x_{n}$ is more complex than $x'_{n}$ if 
$\deg(x_{n}) >\deg\left(x'_{n}\right)$. This definition
of the complexity scales is similar to the organization of the growth rates
of the exponential polynomials as described e.g. in \cite{Levin}. Empirical
(e.g. numerical) studies of simple examples, e.g. of the binary sequences,
show that it captures the intuitive idea that, e.g., $010101010101$ is simpler 
than $010001101111$, and helps to establish a number of beautiful relationships
and a surprisingly rich complexity structure.

It is clear however, that the case of finite discrete sequences over 
finite alphabets is simpler than the case of the infinite sequences 
with real valued elements, as in the case of the spectral sequences. 
Although in some cases it is possible to study the complexity of 
discretized sequences \cite{Muchnik}, for spectral sequences it is 
not clear a priori which discretization scheme should be used.

Using the bootstrapping hierarchy approach leads to a natural scale of
complexity for the spectral sequences. The bootstrapping transformations 
suggested by the use of the Gutzwiller's trace formula, also produce more 
ordered sequences out of less ordered ones, and generate a finite 
complexity degree for a number of systems, such as spectral sequences of 
quantum graphs or the nontrivial zeroes of the Riemann's zeta function. 
This degree is analogous to the polynomial degree of the finite sequences 
as defined in \cite{Arnold1,Arnold2}.

However, an important feature of the finite sequence complexity structures 
established in \cite{Arnold1,Arnold2} via (\ref{arnold}) 
is that they do not generalize in any trivial way when the length of a 
sequence is increased. On the one hand, in many applications, a given 
$N$-letter long sequence over an $M$-letter alphabet may appear as an 
approximation to a longer (e.g. infinite) $N_{1}$-letter long sequence, 
$N_{1}>N$, defined over a larger $M_{1}$-letter alphabet, $M_{1}>M$,
and the task is to characterize the complexity of the entire sequence. 
Hence it is natural to use complexity measures that are stable with 
respect to such completions of shorter sequences by the longer ones.
In the context of studying spectral sequences, complexity index defined 
for a sufficiently long list of elements should stabilize with the 
increase of sequence's length, so that the spectrum as a whole is 
characterized by a single coherent complexity degree, that can be
deduced from sufficiently long finite approximations.

In view of this, it is particularly significant that the degree produced 
by the bootstrapping method proposed above allows a stable characterization 
of the complexity of the whole spectral sequence. Numerical analysis of 
Riemann's zeroes and of the spectra of quantum graphs of different 
topologies shows that the regularity degree obtained on the basis of a 
few hundred levels remains the same when a much larger ($10^{5}-10^{6}$) 
set of levels is considered. 

It is also physically relevant that the expansion (\ref{Gutzw}) is typically
derived with a semiclassical accuracy, so the locations of the peaks
generated by the sum (\ref{Gutzw}) corresponds to the actual $E_{n}$ values 
only approximately. Moreover, even in cases when Gutzwiller's formula is exact, 
as in the case of the quantum graphs \cite{QGT,Nova} and a few other systems 
\cite{Berry3,Gutzw,Selberg,Melrose} the sum (\ref{Gutzw}) cannot in general be 
computed exactly, because only a finite number of the periodic orbits may be 
known and their characteristics (e.g. the factor coefficients $B_{p}$ and the 
actions $S_{p}$) are described with a finite accuracy. Hence the harmonic 
expansions on each level of the hierarchy provide only a ``fuzzy'' description 
of the spectrum. It is therefore important that the regularity degree obtained 
via the bootstrapping method is also stable with respect to the using finite 
approximations to Gutzwiller's sum, so the semiclassical description gives a
correct estimate of the exact hierarchy index of complexity of quantum spectra.

The work was supported in part by the Sloan and Swartz Foundations.

\end{document}